\documentclass[aps,amssymb,twocolumn]{revtex4}
\usepackage{amssymb}
\usepackage{graphicx}
\usepackage{color}
\usepackage{textcomp}
\usepackage{ulem}

\begin{document}
\title{Mass spectra of neutral mesons $K_0,\ \pi_0,\ \eta,\ \eta'$ at finite magnetic field, temperature and baryon chemical potential}
\author{Jie Mei$^1$}
\author{Tao Xia$^2$}
\author{Shijun Mao$^1$}%
\email{maoshijun@mail.xjtu.edu.cn}
\affiliation{School of Physics, Xi'an Jiaotong University, Xi'an, Shaanxi 710049, China$^1$\\
College of Advanced Interdisciplinary Studies, National University of Defense Technology, Changsha, Hunan 410073, China$^2$}
\begin{abstract}
The mass spectra of neutral mesons $K_0, \pi_0, \eta, \eta'$ on temperature-quark chemical potential $(T-\mu)$ plane in the presence of a constant magnetic field is investigated in the $SU(3)$ NJL model. As a Goldstone boson of chiral symmetry breaking, the mass of $K_0$ meson increases with temperature and/or quark chemical potential, and we observe two kinds of mass jumps of $K_0$ meson in media, which is induced by the mass jump of constituent quarks and the magnetic field, respectively. Due to the breaking of isospin symmetry between $u$ and $d$ quarks in magnetic fields, the mixing of $\pi_0-\eta- \eta'$ mesons occurs and this leads to rich structures of their mass spectra. For instance, $\pi_0$ mass is influenced by the strange quark. There appear the change of increase ratio of $\pi_0$ mass at high $\mu$ and vanishing $T$ and the $\pi_0$ mass jump crossing over the threshold of two times of strange quark mass at finite $T$ and $\mu$. The mass ordering of $\pi_0, \ \eta,\ \eta'$ mesons varies in media, due to their mass jumps, which are induced by the mass jump of constituent quarks or the magnetic field.	
\end{abstract}
\date{\today}
\maketitle

\section{Introduction}	
The study of hadron properties in QCD media is important for our understanding of strong interaction matter, due to its close relation to QCD phase structure and relativistic heavy ion collision. The chiral symmetry breaking leads to the rich meson and baryon spectra, and the $U_A(1)$ anomaly explains the non-degeneracy of $\eta$ and $\eta'$ mesons~\cite{eta}. The mass shift of hadrons will enhance or reduce their thermal production in relativistic heavy ion collisions, such as the kaon yields and ratios~\cite{k1,k2,k3,nmeson6}.

It is widely believed that the strongest magnetic field in nature may be generated in the initial stage of relativistic heavy ion collisions. The initial magnitude of the field can reach $eB\sim (1-100)m_\pi^2$ in collisions at the Relativistic Heavy Ion Collider and the Large Hadron Collider~\cite{b0,b1,b2,b3,b4}, where $e$ is the electron charge and $m_\pi$ the pion mass in vacuum. Recent years, magnetic field effects on the hadrons attract much attention. As the Goldstone bosons of the chiral (isospin) symmetry breaking, the properties of neutral (charged) pions at finite magnetic field, temperature and density are widely investigated~\cite{c1,c3,hadron1,hadron2,qm1,qm2,sigma1,sigma2,sigma3,sigma4,l1,l2,l3,l4,lqcd5,lqcd6,ding2008.00493,njl2,meson,mfir,ritus5,ritus6,mao1,mao11,mao2,wang,coppola,phi,liuhao3,he,maocharge,maopion,yulang2010.05716,q1,q2,q3,q4,huangamm1,huangamm2,q5,q6,q7,q8,q9,q10}. Another interesting issue is the charged rho meson, which is related to the electromagnetic superconductivity of
the QCD vacuum~\cite{Chernodub:2010qx,Chernodub:2011mc,Callebaut:2011uc,Ammon:2011je,Cai:2013pda,Frasca:2013kka,Andreichikov:2013zba,Liu:2014uwa,Liu:2015pna,Liu:2016vuw,Kawaguchi:2015gpt,Ghosh:2016evc,Ghosh:2017rjo,l1,Luschevskaya:2014mna,Luschevskaya:2015bea,lqcd5,Ding:2020jui,Ghosh}. Furthermore, there are some other works involving $K,$ $\eta$, $\eta'$ and $\phi$ mesons~\cite{su3meson1,q6,su3meson3,su3meson4,houphi}, heavy mesons~\cite{Marasinghe:2011bt,Machado:2013rta,Alford:2013jva,Machado:2013yaa,Cho:2014exa,Cho:2014loa,Dudal:2014jfa,Bonati:2015dka,Gubler:2015qok,Yoshida:2016xgm,Reddy:2017pqp,CS:2018mag} and baryons~\cite{Tiburzi:2008ma,Andreichikov:2013pga,Tiburzi:2014zva,Haber:2014zba,He:2016oqk,Deshmukh:2017ciw,Yakhshiev:2019gvb} in the magnetic field.

In the previous study, the effect of magnetic field on $K,$ $\eta$ and $\eta'$ mesons are mostly considered in vacuum, with vanishing temperature and density~\cite{su3meson1,q6,su3meson3,su3meson4}. Our current work focuses on the mass spectra of neutral mesons $K_0,\ {\bar K}_0$, $\pi_0$, $\eta$ and $\eta'$ mesons at finite magnetic field, temperature and density, which are related to the restoration of chiral symmetry and $U_A(1)$ anomaly. Having the magnetic field of strength compatible with the strong interaction, such as $eB\sim m^2_\pi$, the quark structure of hadrons should be taken into account. We apply the three flavor Nambu-Jona-Lasinio (NJL) model at quark level~\cite{njll2,njl1,njl3,njl4}, where quarks are treated in mean field level and mesons are the quantum fluctuations constructed from the quark bubble. The electromagnetic interaction of the charged constituent quarks leads to a sensitive dependence of the neutral meson properties on the external electromagnetic fields, for instance, the meson mass jump induced by dimension reduction of the constituent quarks\cite{mao1,mao11,mao2,maocharge,maopion,yulang2010.05716,q2,q3}, and the $\pi_0-\eta-\eta'$ mixing due to the breaking of isospin symmetry between $u$ and $d$ quarks\cite{su3meson4,klevansky1}, in magnetic fields.

The rest paper is arranged as follows. We introduce the magnetized $SU(3)$ NJL model, and derive the formula for quarks and neutral mesons mass spectra in Sec.\ref{form}. The numerical results and analysis of neutral meson masses at finite magnetic field, temperature and baryon chemical potential are  presented in Sec.\ref{numerical}. The summary and outlook are in Sec.\ref{sum}.

\section{Formalism}
\label{form}
The three-flavor NJL model under external magnetic field is defined through the Lagrangian density,
\begin{eqnarray}
\mathcal{L}&=&\bar{\psi}\left(i\gamma^{\mu}D_{\mu}-\hat{m}_0\right)\psi+\mathcal{L}_S+\mathcal{L}_{KMT},\\
\mathcal{L}_S&=&G\sum_{\alpha=0}^{8}\left[(\bar{\psi}\lambda_{\alpha}\psi)^2+(\bar{\psi}i\gamma_5\lambda_{\alpha}\psi)^2\right],\nonumber \\
\mathcal{L}_{KMT}&=&-K\left[\text{det}\bar{\psi}(1+\gamma_5)\psi+\text{det}\bar{\psi}(1-\gamma_5)\psi \right].\nonumber
\label{lagrangian}
\end{eqnarray}
The covariant derivative $D_\mu=\partial_\mu-i Q A_\mu$ couples quarks with electric charge $Q=\text{diag}(Q_u,Q_d,Q_s)=\text{diag}(2/3 e,-1/3 e,-1/3 e)$ to a gauge field ${\bf B}=\nabla\times{\bf A}$. Here, we consider magnetic field in $z$ direction by setting $A_\mu=(0,0,x B,0)$ in Landau gauge. $\hat{m}_0=\text{diag}(m^u_0,m_0^d,m_0^s)$ is the current quark mass matrix in flavor space. The four-fermion interaction $\mathcal{L}_S$ represents the interaction in scalar and pseudo-scalar channels, with Gell-Mann matrices $\lambda_{\alpha},\ \alpha=1,2,...,8$ and $\lambda_0=\sqrt{2/3} \mathbf{I}$. The six-fermion interaction or Kobayashi-Maskawa-'t Hooft term $\mathcal{L}_{KMT}$ is related to the $U_A(1)$ anomaly~\cite{tHooft1,tHooft2,tHooft3,tHooft4,tHooft5}.

It is useful to convert the six-fermion interaction into an effective four-fermion interaction in the mean field approximation, and the Lagrangian density can be rewritten as~\cite{klevansky1}
\begin{eqnarray}
\mathcal{L}&=&\bar{\psi}\left(i\gamma^{\mu}D_{\mu}-\hat{m}_0\right)\psi \\
&+&\sum_{a=0}^{8}\left[K_a^-\left(\bar{\psi}\lambda^a\psi\right)^2+K_a^+\left(\bar{\psi}i\gamma_5\lambda^a\psi\right)^2\right]\nonumber\\	&+&K_{30}^-\left(\bar{\psi}\lambda^3\psi\right)\left(\bar{\psi}\lambda^0\psi\right)+K_{30}^+\left(\bar{\psi}i\gamma_5\lambda^3\psi\right)\left(\bar{\psi}i\gamma_5\lambda^0\psi\right)\nonumber\\	&+&K_{03}^-\left(\bar{\psi}\lambda^0\psi\right)\left(\bar{\psi}\lambda^3\psi\right)+K_{03}^+\left(\bar{\psi}i\gamma_5\lambda^0\psi\right)\left(\bar{\psi}i\gamma_5\lambda^3\psi\right)\nonumber\\	&+&K_{80}^-\left(\bar{\psi}\lambda^8\psi\right)\left(\bar{\psi}\lambda^0\psi\right)+K_{80}^+\left(\bar{\psi}i\gamma_5\lambda^8\psi\right)\left(\bar{\psi}i\gamma_5\lambda^0\psi\right)\nonumber\\	&+&K_{08}^-\left(\bar{\psi}\lambda^0\psi\right)\left(\bar{\psi}\lambda^8\psi\right)+K_{08}^+\left(\bar{\psi}i\gamma_5\lambda^0\psi\right)\left(\bar{\psi}i\gamma_5\lambda^8\psi\right)\nonumber\\	&+&K_{83}^-\left(\bar{\psi}\lambda^8\psi\right)\left(\bar{\psi}\lambda^3\psi\right)+K_{83}^+\left(\bar{\psi}i\gamma_5\lambda^8\psi\right)\left(\bar{\psi}i\gamma_5\lambda^3\psi\right)\nonumber\\	&+&K_{38}^-\left(\bar{\psi}\lambda^3\psi\right)\left(\bar{\psi}\lambda^8\psi\right)+K_{38}^+\left(\bar{\psi}i\gamma_5\lambda^3\psi\right)\left(\bar{\psi}i\gamma_5\lambda^8\psi\right)\nonumber,
\label{semilagrangian}
\end{eqnarray}
with the effective coupling constants
\begin{eqnarray}
\label{constants}
&&K_0^\pm=G\pm\frac{1}{3}K\left(\sigma_u+\sigma_d+\sigma_s\right),\\
&&K_1^\pm=K_2^\pm=K_3^\pm=G\mp\frac{1}{2}K\sigma_s,\nonumber\\
&&K_4^\pm=K_5^\pm=G\mp\frac{1}{2}K\sigma_d,\nonumber\\
&&K_6^\pm=K_7^\pm=G\mp\frac{1}{2}K\sigma_u,\nonumber\\
&&K_8^\pm=G\mp\frac{1}{6}K\left(2\sigma_u+2\sigma_d-\sigma_s\right),\nonumber\\
&&K_{03}^\pm=K_{30}^\pm=\pm\frac{1}{2\sqrt{6}}K\left(\sigma_u-\sigma_d\right),\nonumber\\
&&K_{08}^\pm=K_{80}^\pm=\mp\frac{\sqrt{2}}{12}K\left(\sigma_u+\sigma_d-2\sigma_s\right),\nonumber\\
&&K_{38}^\pm=K_{83}^\pm=\mp\frac{1}{2\sqrt{3}}K\left(\sigma_u-\sigma_d\right),	\nonumber
\end{eqnarray}
and chiral condensates
\begin{eqnarray}
\sigma_u=\langle\bar{u}u\rangle, \  \sigma_d=\langle\bar{d}d\rangle, \  \sigma_s=\langle\bar{s}s\rangle.
\end{eqnarray}

At finite temperature $T$, quark chemical potential $\mu=\mu_B/3$ and magnetic field $eB$, the chiral condensates or effective quark masses $m_u=m_0^u-4G\sigma_u+2K\sigma_d\sigma_s$, $m_d=m_0^d-4G\sigma_d+2K\sigma_u\sigma_s$, $m_s=m_0^s-4G\sigma_s+2K\sigma_u\sigma_d$ are determined by minimizing the thermodynamic potential,
\begin{eqnarray}
\partial\Omega_{\text {mf}}/\partial \sigma_i=0,\ i=u,d,s,
\end{eqnarray}
where the thermodynamic potential in mean field level contains the mean field part and quark part
\begin{eqnarray}
\label{omega1}
\Omega_{\text {mf}} &=&2 G(\sigma_u^2+\sigma_d^2+\sigma_s^2)-4K\sigma_u\sigma_d\sigma_s+\Omega_q,\\
\Omega_q &=& -3 \sum_{f=u,d,s}\frac{|Q_f B|}{2\pi}\sum_{l}\alpha_l \int \frac{d p_z}{2\pi} \Bigg[E_f \nonumber\\
&& +T\ln\left(1+e^{- \frac{E_f+\mu}{T}}\right)+T\ln\left(1+e^{- \frac{E_f-\mu}{T}}\right)\Bigg],\nonumber
\end{eqnarray}
with quark energy $E_f=\sqrt{p^2_z+2 l |Q_f B|+m_f^2}$ of flavor $f=u,d,s$, longitudinal momentum $p_z$ and  Landau level $l$, and the degeneracy of Landau levels $\alpha_l=2-\delta_{l0}$.
	
%The ground state is determined by minimizing the thermodynamic potential,
%\begin{eqnarray}
%\partial\Omega_{\text {mf}}/\partial \sigma_i=0,\ i=u,d,s,
%\end{eqnarray}
%which leads to the coupled gap equations for the chiral condensates of quarks.
	
In the NJL model, mesons are treated as quantum fluctuations above the mean field. Through the random phase approximation (RPA) method~\cite{njl1,njll2,njl3,njl4,njl2}, the meson propagator can be expressed in terms of the irreducible polarization function or quark bubble,
\begin{eqnarray}
\Pi_{M'M}^P(k)=i \text{Tr}\left[\Gamma_{M'}^* S(p+\frac{1}{2}k)\Gamma_{M} S(p-\frac{1}{2}k)\right],
\label{pimm}
\end{eqnarray}
with the quark propagator matrix $S=diag(S_u, S_d, S_s)$ in flavor space, the meson vertex
\begin{equation}
\Gamma_M=\left\{
\begin{array}{l}
i\gamma_5\lambda_0,\ \ M=\eta_0\\
i\gamma_5\lambda_3,\ \ M=\pi_0\\
i\gamma_5(\lambda_6\pm i\lambda_7)/\sqrt{2},\ \ M=K_0,\bar{K}_0\ \ \\
i\gamma_5\lambda_8,\ \ M=\eta_8
\end{array}	\right. ,
\end{equation}
and the trace Tr in spin, color, flavor and momentum space. For neutral mesons $\pi_0,\ K_0,\ \bar{K}_0,\ \eta,\ \eta'$, the Schwinger phase arising from quark propagators is cancelled, and the meson momentum $k=(k_0,\vec{k})$ itself is conserved.

Let's start from neutral kaon mesons ($K_0,\ {\bar K}_0$), which are not mixing with other mesons. The $K_0$ meson propagator can be written as
\begin{eqnarray}
{\cal M}(k_0,\vec{k})=\frac{2K_{6}^+}{1-2K_{6}^+ \Pi_{K_0 K_0}^P(k_0,\vec{k})},
\end{eqnarray}
and the kaon mass is determined through the pole equation at zero momentum $\vec{k}=\vec{0}$,
\begin{eqnarray}
1-2K_6^+ \Pi_{K_0 K_0}^P(m_{K_0},\vec{0})=0.
\label{kaon}
\end{eqnarray}
Since the $K_0$ meson is charge neutral, it is affected by the external magnetic field only through the constituent quarks. The formula for meson propagator is the same as that without magnetic field except for the consideration of Landau levels in momentum integral. The polarization function is simplified as
\begin{eqnarray}
&\Pi_{K_0 K_0}^P(k_0,\vec{0})=\ \ \ \ \ \ \ \ \ \ \ \ \ \ \ \ \ \ \ \ \ \ \ \ \ \ \ \ \ \ \ \ \ \ \ \ \ \ \ \ \ \ \ \ \ \ \ \ \ \nonumber \\
&\ \ \ \ J_1^{(d)}+J_1^{(s)}+2\left(\left(m_d-m_s
\right)^2-k_0^2\right) J_2^{(ds)}(k_0^2)\ \ \ \ \ \
\label{pikaon}
\end{eqnarray}
with
\begin{eqnarray}
&&J_1^{(f)}=3\!\sum_{l}\alpha_l \frac{|Q_f B|}{2\pi}\!\int \frac{dp_z}{2\pi}\frac{\tanh\frac{E_f+\mu}{2T}+\tanh\frac{E_f-\mu}{2T}}{2E_f} \nonumber\\
&&J_2^{(ds)}(k_0^2)=-3\!\sum_{l}\!\alpha_l \frac{|Q_f B|}{2\pi}\int \frac{dp_z}{2\pi}\frac{1}{8 E_s E_d}\ \ \ \ \nonumber\\
&\times & \Bigg[\frac{1}{E_s+E_d+k_0}\left(\tanh\frac{E_s-\mu}{2T}+\tanh\frac{E_d+\mu}{2T}\right) \nonumber\\
&& +\frac{1}{E_s-E_d+k_0} \left(\tanh\frac{E_d-\mu}{2T}-\tanh\frac{E_s-\mu}{2T} \right) \nonumber\\
&& + \frac{1}{E_s+E_d-k_0} \left(\tanh\frac{E_d-\mu}{2T}+\tanh\frac{E_s+\mu}{2T} \right) \nonumber\\
&& + \frac{1}{E_s-E_d-k_0} \left(\tanh\frac{E_d+\mu}{2T}-\tanh\frac{E_s+\mu}{2T} \right) \nonumber \Bigg].\nonumber
\label{kaonj2}	
\end{eqnarray}

As Goldstone boson of chiral symmetry restoration, when the constituent quark mass decreases, the kaon mass will increase. Therefore, accompanied with the chiral symmetry restoration, it is expected to have the intersection between the kaon mass and the sum of two constituent quark masses, which defines the Mott transition of kaon meson~\cite{mott1,zhuang,mott2,mott3}. Based on the polarization function (\ref{pikaon}), with $p_z=0$ and lowest Landau level $l=0$, the integral term $\frac{1}{E_s+E_d-k_0}$ diverges when $k_0=m_d+m_s$. This infrared divergence will lead to the mass jump of kaon meson at the Mott transition, as shown in Fig.\ref{fig:kT} and Fig.\ref{fig:kCEP}.

By interchanging two constituent quarks $E_d \leftrightarrow E_s$, we obtain the polarization function of $\bar{K}_0$ meson. When the baryon chemical potential is zero, $K_0$ and $\bar{K}_0$ mesons share the same mass. For finite baryon chemical potential, they show mass splitting.

Since the magnetic field breaks the isospin symmetry for $u$ and $d$ quarks, the coupling constant $K_{03}$ and $K_{38}$ are no longer zero. The flavor mixing of $\pi_0-\eta-\eta'$ happens. Therefore, the meson propagator can be constructed in a matrix form with the RPA method,
\begin{eqnarray}
{\cal M}=2K^+ (1-2\Pi^P K^+)^{-1},
\end{eqnarray}
where coupling constant $K^+$ and polarization function $\Pi^P$ are $3\times3$ matrices
\begin{eqnarray}
K^+={
    	\left( \begin{array}{ccc}
    		K_0^+ & K_{03}^+ & K_{08}^+ \\
    		K_{30}^+ & K_3^+ & K_{38}^+ \\
    		K_{80}^+ & K_{83}^+ & K_8^+
    	\end{array} \right)},\\
    \Pi^P={
    	\left( \begin{array}{ccc}
    		\Pi_0^P & \Pi_{03}^P & \Pi_{08}^P \\
    		\Pi_{30}^P & \Pi_3^P & \Pi_{38}^P \\
    		\Pi_{80}^P & \Pi_{83}^P & \Pi_8^P
    	\end{array} \right)}.\
\end{eqnarray}
The matrix elements of coupling constant $K^+$ are written in Eq.(\ref{constants}), and the elements of polarization function $\Pi^P$ are defined in Eq.(\ref{pimm}) with index $3,0,8$ denoting $\pi_0, \eta_0, \eta_8$, respectively. For convenience, we sometimes omit the argument $(k_0,\vec{k})$ in the polarization function and meson propagator.

We can obtain $\pi_0,\ \eta,\ \eta'$ meson masses by solving the equation at $\vec{k}=\vec{0}$,
\begin{eqnarray}
\text{det}\left[{\cal M}^{-1}(k_0,\vec{0}) \right]=0.
\end{eqnarray}
The inverse of meson propagator matrix ${\cal M}$ can be simplified as
    \begin{eqnarray}
    	&&{\cal M}^{-1}=\frac{1}{2\text{det}K^+}{
    		\left( \begin{array}{ccc}
    			\mathcal{A} & \mathcal{B} & \mathcal{C} \\
    			\mathcal{B} & \mathcal{D} & \mathcal{E} \\
    			\mathcal{C} & \mathcal{E} & \mathcal{F}
    		\end{array} \right)},\ \\
       	&&\mathcal{A}=\left(K_3^+ K_8^+-K_{38}^{+2}\right) -2\Pi^P_0 \text{det}K^+, \nonumber\\
    	&&\mathcal{B}=\left(K_{38}^+ K_{08}^+-K_{8}^{+}K_{03}^{+}\right) -2\Pi^P_{03} \text{det}K^+, \nonumber\\
    	&&\mathcal{C}=\left(K_{03}^+ K_{38}^+-K_{3}^{+}K_{08}^{+}\right) -2\Pi^P_{08} \text{det}K^+, \nonumber\\
    	&&\mathcal{D}=\left(K_0^+ K_8^+-K_{08}^{+2}\right) -2\Pi^P_3 \text{det}K^+, \nonumber\\
    	&&\mathcal{E}=\left(K_{03}^+ K_{08}^+-K_{0}^{+}K_{38}^{+}\right) -2\Pi^P_{38} \text{det}K^+, \nonumber\\
    	&&\mathcal{F}=\left(K_3^+ K_0^+-K_{03}^{+2}\right) -2\Pi^P_8 \text{det}K^+ .\nonumber
    \end{eqnarray}
with
    \begin{eqnarray}
    \label{polepieta}
    	&&\Pi^P_0=\frac{2}{3}\left(\Pi^P_{uu}+\Pi^P_{dd}+\Pi^P_{ss}\right),\\
    	&&\Pi^P_3=\Pi^P_{uu}+\Pi^P_{dd},\nonumber\\
    	&&\Pi^P_8=\frac{1}{3}\left(\Pi^P_{uu}+\Pi^P_{dd}+4\Pi^P_{ss}\right),\nonumber\\
    	&&\Pi^P_{03}=\Pi^P_{30}=\frac{\sqrt{6}}{3}\left(\Pi^P_{uu}-\Pi^P_{dd}\right),\nonumber\\
    	&&\Pi^P_{08}=\Pi^P_{80}=\frac{\sqrt{2}}{3}\left(\Pi^P_{uu}+\Pi^P_{dd}-2\Pi^P_{ss}\right),\nonumber\\
    	&&\Pi^P_{38}=\Pi^P_{83}=\frac{\sqrt{3}}{3}\left(\Pi^P_{uu}-\Pi^P_{dd}\right),\nonumber
    \end{eqnarray}
and
    \begin{eqnarray}
    	\Pi^P_{ff}(k_0^2)&=&J_1^{(f)}-k^2_0 J_2^{(ff)}(k_0^2),\\
        	 J_2^{(ff)}(k_0^2)&=&-3\!\sum_{l}\!\alpha_l \frac{|Q_f B|}{2\pi}\!\!\int \frac{dp_z}{2\pi}\frac{1}{2E_f\!\left(\!4E_f^2-k_0^2\right)}\nonumber \\
    	&&\times \left(\tanh\frac{E_f+\mu}{2T}+\tanh\frac{E_f-\mu}{2T}\right).
    \end{eqnarray}

Similar as discussed in the kaon meson, each matrix element in polarization function matrix Eq.(\ref{polepieta}) shows infrared divergence at $p_z=0$ and $k_0=2\sqrt{m_f^2+2l|Q_f B|}$ with $f=u,d,s,\ l=0,1,2,...$, due to the integral term $\frac{1}{4E_f^2-k_0^2}$. Because of the $\pi_0-\eta-\eta'$ meson mixing, such infrared divergence will lead to several mass jumps for $\pi_0,\ \eta,\ \eta'$ mesons, as shown in Fig.\ref{fig:peep20} and Fig.\ref{fig:peepCEP}.%Traditionally Mott transition is referred as the transition of a meson from a stable particle to a resonant state. For convenience, in this paper, all the meson mass jumps that follow this machanism are called "Mott transitions".\\

Because of the contact interaction in NJL model, the ultraviolet divergence can't be eliminated through renormalization, and a proper regularization scheme is needed. In our work, we apply the Pauli-Villars regularization~\cite{mao1,mao11,mao2,maopion}, which is gauge invariant and can guarantee the law of causality at finite magnetic field. By fitting the physical quantities, pion mass $m_{\pi}=138\text{MeV}$, pion decay constant $f_{\pi}=93\text{MeV}$, kaon mass $m_K=495.7\text{MeV}$, $\eta'$ meson mass $m_{\eta\prime}=957.5\text{MeV}$ in vacuum, we obtain the parameters $m_0^{u}=m_0^{d}=5.5\text{MeV}$, $m_0^s=154.7\text{MeV}$, $G\Lambda^2=3.627$, $K\Lambda^5=92.835$, $\Lambda=1101\text{MeV}$. In the following numerical calculations, we fix magnetic field $eB=20m^2_{\pi}$.
%\begin{eqnarray}
%    	&&m_{\pi}=138\text{MeV},\ f_{\pi}=93\text{MeV},\nonumber\\
%    	&&m_K=495.7\text{MeV},\ m_{\eta\prime}=957.5\text{MeV},\nonumber
%    \end{eqnarray}

%\begin{eqnarray}
%    	&& m_{u}=m_{d}=5.5\text{MeV},\ m_s=154.7\text{MeV},\nonumber\\
%    	&&G\Lambda^2=3.627,\ K\Lambda^5=92.835,\ \Lambda=1101\text{MeV},\nonumber
%    \end{eqnarray}
\section{Results and analysis}
\label{numerical}
%%%%%%%%%%%%%%%%%%%%%%%%%%%%%%%%%%%%%%%%%%%%%%%%%%%%%%%%%%%%%%%%%%%
\begin{figure}[htbp!]
\includegraphics[width=\columnwidth]{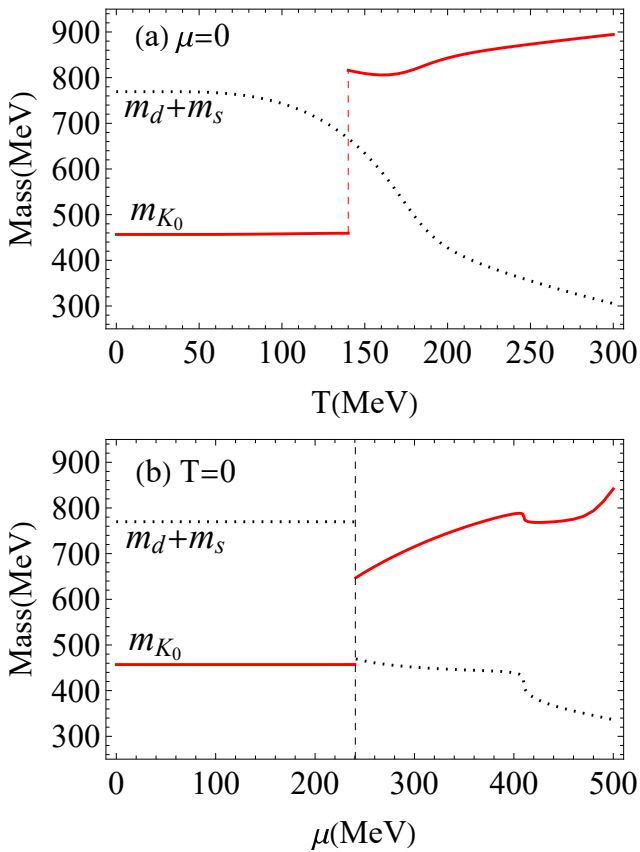}% Here is how to import EPS art
\caption{\label{fig:kT}$K_0$ meson mass $m_{K_0}$ (red solid lines) and quark mass $m_d+m_s$ (black dotted lines) at finite magnetic field $eB=20m_{\pi}^2$ with vanishing quark chemical potential $\mu=0$ in panel (a) and with vanishing temperature $T=0$ in panel (b). The vertical dashed lines are used to denote the sudden mass jumps for quarks (in black) and $K_0$ meson (in red).}
\end{figure}
%%%%%%%%%%%%%%%%%%%%%%%%%%%%%%%%%%%%%%%%%%%%%%%%%%%%%%%%%%%%%%%%%%%

Fig.\ref{fig:kT}(a) plots the mass of $K_0$ meson $m_{K_0}$ and the mass of the two constituent quarks $m_d+m_s$ as functions of temperature $T$ at finite magnetic field $eB=20m_{\pi}^2$ and vanishing quark chemical potential $\mu=0$. The chiral symmetry restoration is a smooth crossover with $\mu=0$, and the quark mass decreases continuously with the increase of temperature. As the Goldstone boson, the mass of $K_0$ meson monotonically increases with temperature, and a sudden mass jump happens at the Mott transition temperature $T_{\text {Mott}}^{K_0}=140.0$MeV, where the $K_0$ mass jumps from the bound state $m_{K_0}<m_d+m_s$ to the resonant state $m_{K_0}>m_d+m_s$. This mass jump is caused by the finite magnetic field, which leads to the dimension reduction of constituent quarks and the infrared divergence of the meson polarization function, as analyzed in Eq.(\ref{pikaon}). With $T>T_{\text {Mott}}^{K_0}$, the mass of $K_0$ meson slightly decreases with temperature and then turns to increase.

Fig.\ref{fig:kT}(b) plots the mass of $K_0$ meson $m_{K_0}$ and the mass of the two constituent quarks $m_d+m_s$ as functions of quark chemical potential $\mu$ at finite magnetic field $eB=20m_{\pi}^2$ and vanishing temperature $T=0$. The chiral symmetry restoration is a first order phase transition with increasing quark chemical potential and $T=0$, and the quark mass shows a jump at $\mu=240.1 {\text {MeV}}$. This mass jump of constituent quarks also leads to the mass jump of $K_0$ meson. Since the $K_0$ mass jumps from $m_{K_0}<m_d+m_s$ to $m_{K_0}>m_d+m_s$, and satisfies the condition $m_{K_0}>2\mu$, this is also a Mott transition. In the chiral breaking phase with $\mu<240.1 {\text {MeV}}$, the quark mass keeps constant. After the chiral restoration phase transition, the quark mass decreases with different ratio. With $240.1 {\text {MeV}}<\mu< 410.2 {\text {MeV}}$, it decreases slowly, and with $\mu\geq 410.2 \ {\text {MeV}}$, the quark mass decreases faster. Note that at $\mu= 410.2 {\text {MeV}}$, the strange quark mass decreases abruptly. For $K_0$ meson, the mass keeps constant in the chiral breaking phase with $\mu<240.1 {\text {MeV}}$, and increases in the chiral restoration phase with $\mu>240.1 {\text {MeV}}$. However, accompanied with the sudden change of the decrease ratio of quark mass, a fast decrease of $K_0$ mass happens around $\mu=410.2 {\text {MeV}}$.

%%%%%%%%%%%%%%%%%%%%%%%%%%%%%%%%%%%%%%%%%%%%%%%%%%%%%%%%%%%%%%%%%%%
\begin{figure}[htbp!]
\includegraphics[width=\columnwidth]{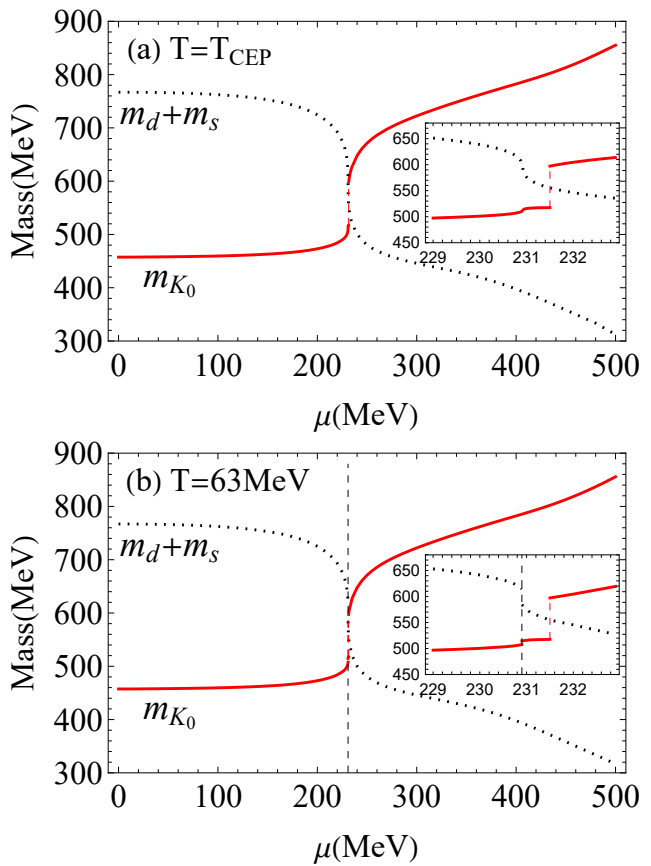}% Here is how to import EPS art
\caption{\label{fig:kCEP}$K_0$ meson mass $m_{K_0}$ (red solid lines) and quark mass $m_d+m_s$ (black dotted lines) at finite magnetic field  $eB=20m_{\pi}^2$ around the critical end point. In panel (a), we fixed temperature at the CEP. Panel (b) is a case of the first order chiral phase transition near CEP. The vertical dashed lines are used to denote the sudden mass jumps for quarks (in black) and $K_0$ meson (in red).}
\end{figure}
%%%%%%%%%%%%%%%%%%%%%%%%%%%%%%%%%%%%%%%%%%%%%%%%%%%%%%%%%%%%%%%%%%

At finite temperature and quark chemical potential, there exists a critical end point (CEP) of chiral symmetry restoration, which connects the crossover and the first order phase transition, and is located at $(T_{\text {CEP}}, \mu_{\text {CEP}})=(63.3 {\text {MeV}}, 230.9 {\text {MeV}})$ with $eB=20m_{\pi}^2$. Fig.\ref{fig:kCEP} depicts the $K_0$ mass $m_{K_0}$ and the mass of two constituent quarks $m_d+m_s$ around CEP. In Fig.\ref{fig:kCEP}(a), we fix temperature at CEP $T=T_{\text {CEP}}$. The quark mass decreases with quark chemical potential, with the fastest change $\frac{d m_q}{d\mu} \rightarrow -\infty$ at $\mu=\mu_{\text {CEP}}$. The $K_0$ mass increases with quark chemical potential, with the fastest change $\frac{d m_{K_0}}{d\mu} \rightarrow \infty$ at $\mu=\mu_{\text {CEP}}$, too. At $\mu>\mu_{\text {CEP}}$, the increase ratio of $K_0$ mass becomes finite, and the Mott transition with $K_0$ mass jump happens at $\mu_{\text {Mott}}^{K_0}=231.5 {\text {MeV}}>\mu_{\text {CEP}}$. At $\mu>\mu_{\text {Mott}}^{K_0}$, $K_0$ mass continues to increase. Around the CEP with $T>T_{\text {CEP}}$, the chiral restoration is a smooth crossover, and the behavior of quark mass and $K_0$ mass is similar as in Fig.\ref{fig:kCEP}(a), but with finite mass change ratio. On the other side, as shown in Fig.\ref{fig:kCEP}(b) with $T<T_{\text {CEP}}$, the chiral restoration is a first order phase transition. The quark mass keeps decreasing with quark chemical potential, associated with a mass jump at $\mu=230.9{\text {MeV}}$. Accordingly, $K_0$ mass keeps increasing, and we observe two mass jumps, caused by the quark mass jump at $\mu=230.9{\text {MeV}}$ and the magnetic field at $\mu=231.5{\text {MeV}}$, respectively. Comparing with Fig.\ref{fig:kT}, the non-monotonical behavior of $m_{K_0}$ disappears around CEP.
	
%%%%%%%%%%%%%%%%%%%%%%%%%%%%%%%%%%%%%%%%%%%%%%%%%%%%%%%%%%%%%%%%%%%
\begin{figure}[htbp!]
\includegraphics[width=\columnwidth]{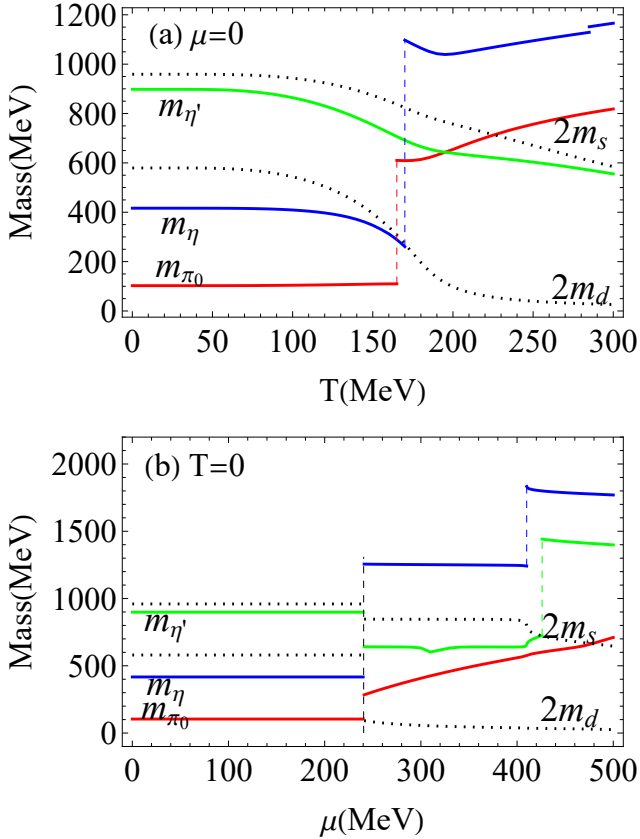}% Here is how to import EPS art
\caption{\label{fig:peep20}Panel (a) is temperature dependence of $\pi_0$ (red solid lines), $\eta$ (blue solid lines), $\eta'$ (green solid lines) masses and quark masses (black dotted lines) with $\mu=0$ and $eB=20m_{\pi}^2$. Panel (b) is quark chemical potential dependence of them at $T=0$ and $eB=20m_{\pi}^2$. The vertical dashed lines are used to denote the sudden mass jumps for quarks and mesons.}
\end{figure}
%%%%%%%%%%%%%%%%%%%%%%%%%%%%%%%%%%%%%%%%%%%%%%%%%%%%%%%%%%%%%%%%%%%

\begin{figure}[htbp!]
\includegraphics[width=\columnwidth]{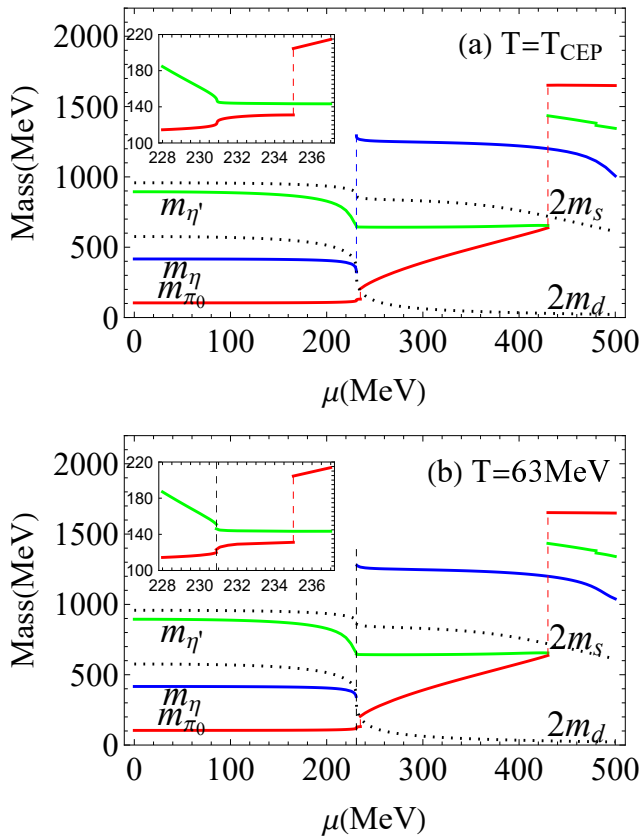}% Here is how to import EPS art
\caption{\label{fig:peepCEP}Masses of $\pi_0$ (red solid lines), $\eta$ (blue solid lines), $\eta'$ (green solid lines) and quarks (black dotted lines) with $eB=20m_{\pi}^2$ around the critical end point. In panel (a), we fixed temperature at the CEP. Panel (b) is a case of the first order chiral phase transition near CEP. The vertical dashed lines are used to denote the sudden mass jumps for quarks and mesons. In the small window, we plot $m_{\pi_0}$ in red and $m_{\eta'}-500\ {\text {MeV}}$ in Green.}
\end{figure}

Fig.\ref{fig:peep20} and Fig.\ref{fig:peepCEP} show the masses of $\pi_0, \ \eta,\ \eta'$ mesons and the constituent quark masses $2m_d,\ 2m_s$ at finite magnetic field, temperature and quark chemical potential. Since $u$-quark mass is very close to (a little bit larger than) $d$-quark, to make the figure clear, we omit its lines. With finite magnetic field, the mixing of $\pi_0-\eta- \eta'$ mesons leads to rich structures of their mass spectra.

In Fig.\ref{fig:peep20}(a), with the chiral crossover or continuous decreasing of the quark mass at finite temperature, $\pi_0$ mass increases slowly in the low temperature region $T<165.0{\text {MeV}}$ and a mass jump from $m_{\pi_0}<2m_d$ to $2m_d<m_{\pi_0}<2m_s$ happens at $T=165.0{\text {MeV}}$. After that, $\pi_0$ mass slightly decreases and then turns to increase with temperature. When the $\pi_0$ mass crosses over the two times of the strange quark mass at high temperature, it increases smoothly, which indicates that the $\pi_0$ meson does not contain the strange quark, even with the $\pi_0-\eta-\eta'$ mixing. For $\eta$ meson, which contains the strange quark, its mass decreases with temperature at $T<170.0{\text {MeV}}$. At $T=170.0{\text {MeV}}$, the mass of $\eta$ meson jumps from $m_{\eta}<2m_d$ to $m_{\eta}>2m_s$. After that, the mass of $\eta$ meson firstly decreases and later increases with temperature. At $T=285.0{\text {MeV}}$, another mass jump occurs. $\eta'$ meson is a resonant state in vacuum, which has the mass larger than two times of the $d$-quark. With the increase of temperature, the mass of $\eta'$ meson continuously decreases, and it becomes lower than $\pi_0$ mass at $\mu=195.0{\text {MeV}}$. In the whole temperature region, we observe $2m_d<m_{\eta'}<2m_s$.

In Fig.\ref{fig:peep20}(b), the first order chiral phase transition at $T=0$ and $\mu=240.1{\text {MeV}}$ leads to the mass jumps for quarks, and also causes the mass jumps of $\pi_0, \ \eta,\ \eta'$ mesons. Before the mass jump, the mass of $\pi_0, \ \eta,\ \eta'$ mesons keeps their value in vacuum, respectively. At $\mu>240.1{\text {MeV}}$, $\pi_0$ mass increases with quark chemical potential. However, the increase ratio changes noncontinuously at $\mu=410.2{\text {MeV}}$, where the strange quark mass abruptly decreases, and at $\mu=478.2{\text {MeV}}$, where the $\pi_0$ mass crosses over the two times of strange quark mass. This indicates that the $\pi_0$ meson is influenced by the strange quark, due to the mixing of $\pi_0- \eta- \eta'$ mesons. For $\eta$ meson, its mass jumps up from $m_{\eta}<2m_d$ to $m_{\eta}>2m_s$ at $\mu=240.1{\text {MeV}}$, and after that, it decreases slightly. At $\mu=410.2{\text {MeV}}$, where the strange quark mass abruptly decreases, another mass jump of $\eta$ meson happens, and then $m_{\eta}$ goes down with quark chemical potential. $\eta'$ meson is in resonant state and its mass jumps down associated with the quark mass jump at $\mu=240.1 {\text {MeV}}$. At $\mu=426.0 {\text {MeV}}$, where the $\eta'$ mass crosses over the two times of strange quark mass, another mass jump from $m_{\eta'}<2m_s$ to $m_{\eta'}>2m_s$ happens. In the region $240.1 {\text {MeV}}<\mu<426.0 {\text {MeV}}$, $\eta'$ mass firstly decreases with quark chemical potential and then increases, with a local minimum at $\mu=310.6 {\text {MeV}}$ and the maximum increase ratio at $\mu=410.2 {\text {MeV}}$, where the strange quark mass decreases abruptly. At $\mu > 426.0 {\text {MeV}}$, $\eta'$ mass goes down with quark chemical potential. %The mass ordering of $\pi_0, \ \eta,\ \eta'$ mesons does not change at vanishing temperature, with $m_{\pi_0}<m_{\eta}<m_{\eta'}$ in the whole consider quark chemical potential region.
	
%%%%%%%%%%%%%%%%%%%%%%%%%%%%%%%%%%%%%%%%%%%%%%%%%%%%%%%%%%%%%%%%%%%
%%%%%%%%%%%%%%%%%%%%%%%%%%%%%%%%%%%%%%%%%%%%%%%%%%%%%%%%%%%%%%%%%%%

Comparing Fig.\ref{fig:peep20}(a) and Fig.\ref{fig:peep20}(b), the structure of meson mass spectra behaves very different, which demonstrates that the temperature and quark chemical potential have different effect on $\pi_0, \ \eta,\ \eta'$ meson masses, respectively. Fig.\ref{fig:peepCEP} depicts the situation near CEP, where panel (a) is at CEP and panel (b) represents an example of the first order chiral phase transition near CEP. To clearly show the near-CEP behavior of $\pi_0$ and $\eta'$ at the same time, in the small window we plot $m_{\pi_0}$ in red and $m_{\eta'}-500\ {\text {MeV}}$ in Green.

Fig.\ref{fig:peepCEP}(a) displays the masses of $\pi_0, \ \eta,\ \eta'$ mesons and the quark masses $2m_d,\ 2m_s$ at finite quark chemical potential with fixed magnetic field $eB=20m_{\pi}^2$ and temperature $T=T_{\text {CEP}}$. $\pi_0$ mass monotonically increases with quark chemical potential. The fastest change $\frac{dm_{\pi_0}}{d\mu} \rightarrow \infty$ is at $\mu=\mu_{\text {CEP}}$, the first mass jump from $m_{\pi_0}<2m_d$ to $2m_d<m_{\pi_0}<2m_s$ happens at $\mu=235.0{\text {MeV}} >\mu_{\text {CEP}}$, and another mass jump from $m_{\pi_0}<2m_s$ to $m_{\pi_0}>2m_s$ happens at $\mu=430.0{\text {MeV}}$. The mass of $\eta$ meson decreases with the quark chemical potential, with the fastest change $\frac{dm_{\eta}}{d\mu} \rightarrow -\infty$ at $\mu=\mu_{\text {CEP}}$ and a mass jump from $m_{\eta}<2m_d$ to $m_{\eta}>2m_s$ at $\mu=230.9{\text {MeV}}$ $>\mu_{\text {CEP}}$. The mass of $\eta'$ meson also decreases with the quark chemical potential. Since it is in resonant state, the mass decrease ratio changes abruptly at $\mu=\mu_{\text {CEP}}$. At $\mu=430.0{\text {MeV}}$, the mass of $\eta'$ meson becomes degenerate with $\pi_0$ meson, and at the same time, the mass jump from $m_{\eta'}<2m_s$ to $m_{\eta'}>2m_s$ occurs. Another mass jump happens at $\mu=480.1 {\text {MeV}}$. Note that the mass ordering of $\pi_0, \ \eta,\ \eta'$ meson varies, with $m_{\pi_0}<m_{\eta}<m_{\eta'}$ in the region $\mu=230.9{\text {MeV}}$ $>\mu_{\text {CEP}}$, $m_{\pi_0}<m_{\eta'}<m_{\eta}$ in the region $230.9{\text {MeV}}<\mu<430.0{\text {MeV}}$ and $m_{\eta}<m_{\eta'}<m_{\pi_0}$ in the region $\mu > 430.0 {\text {MeV}}$, which are influenced by the meson mass jump and the meson mixing under external magnetic field. For the case of chiral crossover near CEP with $T>T_{\text {CEP}}$, the behavior of $\pi_0, \ \eta,\ \eta'$ meson mass is similar as in Fig.\ref{fig:peepCEP}(a), but the infinite change ratio of the mass of $\pi_0, \ \eta$ meson is replaced by the finite value. For the case of first order chiral phase transition near CEP with $T<T_{\text {CEP}}$, as shown in Fig.\ref{fig:peepCEP}(b), instead of  the infinite change ratio of $\pi_0,\ \eta$ meson masses and the abrupt change of $m_{\eta'}$ decrease ratio, we observe one more mass jumps for the $\pi_0, \ \eta,\ \eta'$ meson, caused by the mass jump of the constituent quarks at $\mu=230.9 {\text {MeV}}$, and other behavior of $\pi_0, \ \eta,\ \eta'$ meson masses looks similar as in Fig.\ref{fig:peepCEP}(a).

\section{summary and outlook}
\label{sum}	
The mass spectra of neutral mesons $K_0, \pi_0, \eta, \eta'$ on temperature-quark chemical potential $(T-\mu)$ plane in the presence of a constant magnetic field is studied in the $SU(3)$ NJL model.

As a Goldstone boson of chiral symmetry breaking, the mass of $K_0$ meson increases with temperature and/or quark chemical potential, and the Mott transition happens associated with $K_0$ mass jump. At vanishing $T$ or $\mu$, there exists non-monotonical behavior of $K_0$ mass. Around the CEP, this behavior disappears, and we observe twice mass jumps of $K_0$ meson, which is induced by the mass jump of constituent quarks and the magnetic field, respectively.

Due to the breaking of isospin symmetry between $u$ and $d$ quarks in magnetic fields, the mixing of $\pi_0-\eta- \eta'$ mesons occurs and this leads to rich structures of their mass spectra. The mass of $\pi_0$ meson increases with temperature and/or quark chemical potential, associated with the mass jump, which is similar with the $K_0$ meson. In addition, $\pi_0$ mass is influenced by the strange quark because of the flavor mixing. There appear the change of increasing ratio of $\pi_0$ mass at high $\mu$ and vanishing $T$ and the $\pi_0$ mass jump crossing over the threshold of two times of strange quark mass at finite $T$ and $\mu$. For $\eta$ meson, its mass mainly decreases with $T$ and $\mu$ in the region without jumps. At vanishing $\mu$, it shows twice mass jumps caused by the magnetic field and non-monotonical behavior appears after the first mass jump. At vanishing $T$, it shows twice mass jumps caused by the mass jump of constituent quarks and the magnetic field, respectively. Around the CEP, we observe one mass jump, and the non-monotonical behavior disappears. For $\eta'$ meson, its mass continuously decreases at $\mu=0$, but at $T=0$, it displays twice mass jumps and non-monotonical behavior between the two mass jumps. Around the CEP, $\eta'$ meson mass jumps at high $\mu$, induced by the constituent quark mass jump or the magnetic field. The mass ordering of $\pi_0, \ \eta,\ \eta'$ meson varies in media.

As a consequence of such mass jumps, some interesting phenomena may result in relativistic heavy ion collisions where a strong magnetic field can be created. For instance, there might be a sudden enhancement or reduction of neutral meson production in media, which will be studied in the future. The investigation of charged mesons, $K^\pm$ and $\pi^\pm$ mesons, and the consideration of inverse magnetic catalysis effect is under progress and will be reported elsewhere.

\noindent {\bf Acknowledgement:}
The work is supported by the NSFC Grant 12275204 and Fundamental Research Funds for the Central Universities.\\

%%%%%%%%%%%%%%%%%%%%%%%%%%%%%%%%%%%%%%%%%%%%%	

\end{document}